\documentclass[12pt]{article}
\usepackage{amssymb}

\newtheorem{proposition}{Proposition}

\def\+{{+\!\!\!+}}

\def\d{\partial}

\def\pmb#1{\setbox0=\hbox{#1}%
\kern.0em\copy0\kern-\wd0
\kern-.04em\copy0\kern-\wd0
\kern.08em\copy0\kern-\wd0
\kern-.04em\raise.0433em\box0 }         

\newcommand{\nc}{\newcommand}
\nc{\beq}{\begin{equation}}
\nc{\eeq}[1]{\label{#1}\end{equation}}
\nc{\ber}{\begin{eqnarray}}
\nc{\eer}[1]{\label{#1}\end{eqnarray}}
\nc{\pek}[1]{\cite{#1}}
\nc{\enr}[1]{(\ref{#1})}
\nc{\kal}[1]{{\cal{#1}}}
\nc{\dott}{\;\cdot\;}

\newcommand{\be}{\begin{equation}}
\newcommand{\ee}{\end{equation}}
\newcommand{\bea}{\begin{eqnarray}}
\newcommand{\eea}{\end{eqnarray}}

\newcommand{\Section}[1]{\section{#1} \setcounter{equation}{0}}
\begin{document}

\begin{center}
                                \hfill   hep-th/0502137\\

\vskip .3in \noindent

\vskip .1in

{\large \bf{Hamiltonian perspective on generalized complex structure}}
\vskip .2in

{\bf Maxim Zabzine}\footnote{e-mail address: m.zabzine@qmul.ac.uk} \\


\vskip .15in

\vskip .15in
{\em School of Mathematical Sciences, Queen Mary, University of London}\\
{\em Mile End Road, London E1 4NS, UK}

\bigskip


 \vskip .1in
\end{center}
\vskip .4in

\begin{center} {\bf ABSTRACT }
\end{center}
\begin{quotation}\noindent
 In this note we clarify the relation between  extended world-sheet
 supersymmetry and generalized complex structure. The analysis is 
 based on the phase space description of a  wide class of sigma models. 
We point out the natural isomorphism between the group of orthogonal automorphisms
 of the Courant bracket and the group of local canonical transformations of 
 the cotangent bundle of the loop space. Indeed this fact explains 
 the natural relation between the world-sheet and the geometry of $T\oplus T^*$. 
 We discuss  D-branes in this perspective. 
\end{quotation}
\vfill
\eject


\section{Introduction}

The concept of generalized complex structure was introduced by Hitchin \cite{Hitchin}
 and studied by Gualtieri in his thesis \cite{Gualtieri}. The generalized complex
 structure and related constructions such as generalized K\"ahler and generalized Calabi-Yau
 structures appear naturally in the context  of geometry of the sum of the tangent and 
 cotangent bundles, $T\oplus T^*$.  At the same time there are  indications that 
 the geometry of $T\oplus T^*$ plays a profound role within modern string theory.  
  Actually before the Hitchin's work \cite{Hitchin} some of the relevant mathematical 
 notions were anticipated in the string literature (e.g.,  the algebraic definition
  of a generalized complex (K\"ahler) geometry
 is discussed in \cite{Kapustin:2000aa}). This note intends to further examine
 the relation between the geometry of $T\oplus T^*$ and string theory.
 
 In particular we want to explore the relation between the generalized complex 
  geometry and extended supersymmetry of world-sheet theories. The objective of
  this note is to clarify and extend some of the results from \cite{Lindstrom:2004iw}. 
  Typically the extended supersymmetry for the low dimensional 
 sigma models is related to complex geometry and this is a model independent statement.
 Let us recall the simple algebraic argument for this fact.
  We start from the ansatz
\beq
 \delta(\epsilon) \Phi^\mu = \epsilon D \Phi^\nu J^\mu_{\,\,\nu},
\eeq{normalsusy}
 where $\Phi$ is a superfield corresponding to the map $X : \Sigma \rightarrow M$. 
 A simple calculation of the algebra gives the following expression
\beq
 [\delta(\epsilon_1), \delta (\epsilon_2)] \Phi^\mu = -2 \epsilon_1 \epsilon_2 \d \Phi^\lambda
 (J^\mu_{\,\,\nu} J^\nu_{\,\,\lambda}) - 2\epsilon_1 \epsilon_2 D\Phi^\lambda D\Phi^\rho
 (J^\nu_{\,\,\lambda,\rho} J^\mu_{\,\,\nu} - J^\nu_{\,\,\rho} J^\mu_{\,\,\lambda,\nu}) .
\eeq{susyalg}
 To reproduce the supersymmetry algebra 
\beq
[\delta(\epsilon_1), \delta (\epsilon_2)] \Phi^\mu = 2 \epsilon_1 \epsilon_2 \d \Phi^\mu
\eeq{stasusy2932}
  $J$ is thus a complex structure. 
 The main idea of this note is try to repeat this simple algebraic argument in phase 
 space $(\Phi, S)$ where $S$ is the momentum conjugated to $\Phi$.
  In the phase space 
 writing down the ansatz for the transformation is equivalent to the choice of symplectic
 structure and the generator for the transformation. Using the most general form for the generator
  for second supersymmetry and the standard (twisted) symplectic structure we arrive to
  the main result of the  paper that
 the  phase space realization of extended supersymmetry 
 is related to generalized complex structure. Unlike \cite{Lindstrom:2004iw} all results
 presented in this note are obtained in a model independent way. 
 Indeed the phase space picture offers
 a natural explanation of the appearance of the geometry of $T\oplus T^*$  
  and it agrees with 
  the recent work   \cite{Alekseev:2004np} where the role of the Courant bracket has been discussed
   in this context.
  
The note is organized as follows. In section \ref{s:ham} we start by reviewing 
 the standard description of the string phase space in terms of cotangent 
 bundle $T^*LM$  of the loop space $LM$. Then we introduce the $N=1$ version 
 of $T^*LM$ and explain the notation. We point out the natural isomorphism between 
 the group of orthogonal automorphism of the Courant bracket and the group of local 
 canonical transformations
 of $T^*LM$ (or its supersymmetric version). In section \ref{s:susy} we explain the 
 relation between extended suspersymmetry and generalized complex geometry. 
 We also explain how the real Dirac structures may arise in this context. 
  In the following section \ref{s:branes} we deal with D-branes in the present context.
 We replace the loop space $LM$ by the interval 
 space $PM$. We define the $N=1$ version of $T^*PM$ and explore the relation 
 to generalized complex submanifolds.
  Finally, in section \ref{s:end} we give a summary of the
paper with a discussion of the open problems and the relation of our result to previous results
in the literature. 
 There are two Appendices at the end of the paper. In the first Appendix we establish our conventions
 for $N=1$ superspace. In the second Appendix the basic facts about 
 $T\oplus T^*$ geometry are stated.

\section{Hamiltonian formalism}
\label{s:ham}

 A wide class of sigma models share the following phase space description 
 (e.g., see \cite{Alekseev:2004np}). 
 For the world-sheet $\Sigma = S^1 \times {\mathbb R}$ 
 the phase space can be identified with  a cotangent bundle $T^*LM$ of the loop space 
 $LM=\{ X: S^1 \rightarrow M\}$. Using local coordinates $X^\mu(\sigma)$ and their conjugate 
 momenta $p_\mu(\sigma)$ 
 the standard symplectic form on $T^*LM$ is given by
\beq
 \omega = \int\limits_{S^1} d\sigma \,\, \delta X^\mu \wedge \delta p_\mu ,
\eeq{sympbos}
 where $\delta$ is de Rham differential on $T^*LM$.
 The symplectic form (\ref{sympbos}) can be twisted by a closed three form $H \in \Omega^3(M)$, $dH=0$
 as follows 
\beq
 \omega = \int\limits_{S^1} d\sigma \,\,(\delta X^\mu \wedge \delta p_\mu + \frac{1}{2} H_{\mu\nu\rho} \d X^\mu 
 \delta X^\nu \wedge \delta X^\rho ) ,
\eeq{symptwist}
 where $\d \equiv \d_\sigma$ is derivative with respect to $\sigma$. 
 For both symplectic structures the following transformation is canonical 
\beq
X^\mu\,\,\rightarrow\,\,X^\mu,\,\,\,\,\,\,\,\,\,\,\,\,\,\,\,\,\,\,\,
 p_\mu\,\,\rightarrow\,\,p_\mu + b_{\mu\nu} \d X^\nu
\eeq{canobtransf}
 associated with a closed two form, $b \in \Omega^2(M)$,  $db=0$. There are also canonical 
 transformations which correspond to $Diff(M)$ when $X$ transforms as a coordinate and 
 $p$ as a section of cotangent bundle $T^*M$. In fact the group of local 
  canonical transformations\footnote{By local canonical transformation we mean
 those canonical transformations when new pair $(\tilde{X},\tilde{p})$ are given as
  local expression in terms of the old one $(X, p)$. For example, in the discussion of T-duality 
 one uses non-local canonical transformations, i.e. $\tilde{X}$ is non-local expression in terms of $X$.}
 for $T^*LM$ is a semidirect product of $Diff(M)$ and $\Omega^2_{closed}(M)$.  
  Therefore we come to the following proposition
  \begin{proposition}
  The group of local canonical transformations on $T^*LM$ is isomorphic to the 
 group of orthogonal  automorphisms of Courant bracket. 
 \end{proposition}
  For the description of orthogonal automorphisms of Courant bracket see
  Appendix B. 

 This construction is supersymmetrized in rather straightforward fashion (see Appendix A for 
 superspace conventions). 
 Let $S^{1,1}$ be a "supercircle" with coordinates $(\sigma, \theta)$. Then the corresponding 
 superloop space  is ${\cal L}M = \{ \Phi :  S^{1,1} \rightarrow M\}$. The phase space 
 is given by the cotangent bundle $\Pi T^*{\cal L}M$ of ${\cal L}M$, however with 
  reversed parity on the fibers. In local coordinates we have a scalar superfield $\Phi(\sigma,\theta)$
 and a conjugate momenta, spinorial superfield  $S_\mu(\sigma,\theta)$ with the following
 expansion
\beq
 \Phi^\mu (\sigma, \theta) = X^\mu(\sigma) + \theta \lambda^\mu(\sigma),\,\,\,\,\,\,\,\,\,\,\,\,\,\,\,\,\,\,\,\,\,\,
 S_\mu (\sigma, \theta)= \rho_\mu(\sigma) + \theta p_\mu (\sigma),
\eeq{superfiel}
 where $\lambda$ and $\rho$ are fermions (their linear combinations can be related to the standard 
 world-sheet fermions $\psi_+$ and $\psi_-$).  $S$ is a section of the pullback $X^*(\Pi T^*M)$
  of the cotangent bundle of $M$, considered as an odd bundle.
 The corresponding symplectic structure
 on $\Pi T^*{\cal L}M$ is 
\beq
 \omega = \int\limits_{S^{1,1}} d\sigma d\theta\,\,(\delta \Phi^\mu \wedge \delta S_\mu -
 \frac{1}{2} H_{\mu\nu\rho} D\Phi^\mu \delta \Phi^\nu \wedge \delta \Phi^\rho ),
\eeq{sympsuper}
 such that the bosonic part of (\ref{sympsuper}) coincides with (\ref{symptwist}). 
  Therefore $C^\infty(\Pi T^*{\cal L}M)$ carries the structure of super-Poisson algebra.
 Again as in the purely bosonic case the group of local canonical transformations of $\Pi T^*{\cal L}M$ is 
 a semidirect product of $Diff(M)$ and $\Omega^2_{closed}(M)$.
 The $b$-transform now is given by
\beq
 \Phi^\mu\,\,\rightarrow\,\,\Phi^\mu,\,\,\,\,\,\,\,\,\,\,\,\,\,\,\,\,\,\,\,
 S_\mu\,\,\rightarrow\,\,S_\mu - b_{\mu\nu} D\Phi^\nu,
\eeq{canonfr;as}
 or in components
\ber
 &&X^\mu\,\,\rightarrow\,\,X^\mu, \,\,\,\,\,\,\,\,\,\, p_\mu\,\,\rightarrow\,\,p_\mu + b_{\mu\nu}\d X^\nu 
 +  b_{\mu\nu,\rho} \lambda^\nu
 \lambda^\rho,\\
&&\lambda^\mu\,\,\rightarrow\,\,\lambda^\mu,\,\,\,\,\,\,\,\,\,\,\,\ \rho_\mu\,\,\rightarrow\,\,\rho_\mu 
 - b_{\mu\nu}\lambda^\nu.
\eer{compbtransf}

\section{Supersymmetry in phase space}
\label{s:susy}

In this section we describe the conditions under which extended supersymmetry can 
 be introduced on $\Pi T^*{\cal L}M$. We start from the case $H=0$. By construction 
 of $\Pi T^*{\cal L}M$ the generator of manifest supersymmetry is given by
\beq
 {\mathbf Q}_1(\epsilon) = - \int\limits_{S^{1,1}} d\sigma d\theta\,\, \epsilon S_\mu Q \Phi^\mu ,
\eeq{manisfetsssu}
 where $Q$ is operator introduced in (\ref{relaon1d}) and $\epsilon$ is an odd parameter. 
 Using (\ref{sympsuper}) we can calculate the Poisson brackets for supersymmetry generators
\beq
 \{ {\mathbf Q}_1(\epsilon), {\mathbf Q}_1(\tilde{\epsilon})\} = {\mathbf P}(2\epsilon\tilde{\epsilon}),
\eeq{susyalg1}
 where $P$ is generator of translations along $\sigma$
\beq
 {\mathbf P}(a) = \int\limits_{S^{1,1}} d\sigma d\theta\,\, a S_\mu \d \Phi^\mu
\eeq{defintranms}
 with $a$ being an even parameter. 

 Next we study when there exists a second supersymmetry. The second supersymmetry should 
 be generated by some ${\mathbf Q}_2(\epsilon)$ such that it satisfies the following brackets
\beq
 \{ {\mathbf Q}_1(\epsilon), {\mathbf Q}_2(\tilde{\epsilon})\} = 0,\,\,\,\,\,\,\,\,\,\,\,\,\,\,
 \{ {\mathbf Q}_2(\epsilon), {\mathbf Q}_2(\tilde{\epsilon})\} = {\mathbf P}(2\epsilon\tilde{\epsilon}).
\eeq{susy2regwj}
 By dimensional arguments there is a unique ansatz 
  for the generator ${\mathbf Q}_2(\epsilon)$ on $\Pi T^*{\cal L}M$ which
  does not involve any dimensionful parameters
\beq
 {\mathbf Q}_2(\epsilon) = - \frac{1}{2} \int\limits_{S^{1,1}} d\sigma d\theta\,\,\epsilon
 ( 2 D\Phi^\rho S_\nu J^\nu_{\,\,\rho} + D\Phi^\nu D\Phi^\rho L_{\nu\rho} + 
 S_\nu S_\rho P^{\nu\rho}). 
\eeq{definchat}
 We can combine $D\Phi$ and $S$ into a single object 
\beq
\Lambda = \left ( \begin{array}{l}
                    D\Phi\\
                      S
\end{array} \right)
\eeq{definels}
 which can be thought of as a section of pullback of  $X^*(\Pi (T\oplus T^*))$. The tensors in (\ref{definchat})
 can be combined into a single object\footnote{To relate to other notation
  (e.g., in \cite{Lindstrom:2004iw}), the whole problem is
 invariant under the change $J \rightarrow -J$.}
\beq
 {\cal J} = \left ( \begin{array}{ll}
                    - J & P\\
                     L &  J^t
\end{array} \right),
\eeq{definstardo}
 which is understood now as ${\cal J}: T \oplus T^* \rightarrow T\oplus T^*$.
 With this new notation we can rewrite (\ref{definchat}) as follows
\beq
  {\mathbf Q}_2(\epsilon) = - \frac{1}{2} \int\limits_{S^{1,1}} d\sigma d\theta\,\,\epsilon \langle
 \Lambda, {\cal J} \Lambda \rangle.
\eeq{regska;}
 If the generators ${\mathbf Q}_1(\epsilon)$ and ${\mathbf Q}_2(\epsilon)$ satisfy the 
  algebra (\ref{susyalg1}) and (\ref{susy2regwj}) then we say that there is $N=2$ supersymmetry.
  The following proposition tells us when there exists $N=2$ supersymmetry.
 \begin{proposition}
  $\Pi T^*{\cal L}M$ admits $N=2$ supersymmetry if and only
  if $M$ is a generalized complex manifold.
 \end{proposition}
{\it Proof:}  We have to impose the algebra (\ref{susy2regwj}) on ${\mathbf Q}_2(\epsilon)$.
 The calculation of the second bracket is lengthy but straightforward. The coordinate 
 expressions coincide with those given in \cite{Lindstrom:2004iw}. Therefore we give only the final 
 result of the calculation. Thus the algebra (\ref{susy2regwj}) satisfied if and only if 
\beq 
{\cal J}^2=-1_{2d},\,\,\,\,\,\,\,\,\,\,\,
 \Pi_{\mp} [\Pi_{\pm}(X+\eta), \Pi_{\pm}(Y+\eta)]_c=0,
 \eeq{definsgencomp}
 where $\Pi_\pm=\frac{1}{2}(1_{2d} \pm i {\cal J})$. Thus (\ref{definsgencomp}) together with the fact that 
  ${\cal J}$ (see (\ref{definstardo})) respects  the natural pairing (${\cal J}^t {\cal I} = - {\cal I} {\cal J}$)
 implies that ${\cal J}$ is a generalized complex structure.  $\Pi_\pm$ project to two 
  maximally isotropic involutive subbundles $L$ and $\bar{L}$ such that
$(T\oplus T^*)\otimes {\mathbb C} = L \oplus \bar{L}$.
  Thus we have shown that $\Pi T^*{\cal L}M$ admits $N=2$ supersymmetry if and only
  if $M$ is a generalized complex manifold. Our derivation is algebraic in nature 
   and does not depend on the details of  the model.  
 
 The canonical transformations of $\Pi T^*{\cal L}M$ cannot change any brackets.
  Thus the canonical transformation corresponding to a b-transform (\ref{canonfr;as})
\beq
\left ( \begin{array}{l}
                    D\Phi\\
                      S
\end{array} \right)\,\,\,\rightarrow\,\,\,
\left ( \begin{array}{ll}
                    \,\,\,\,\,1 & 0\\
                   -b & 1
\end{array} \right)
\left ( \begin{array}{l}
                    D\Phi\\
                      S
\end{array} \right)
\eeq{howbtrahsdl}
 induces the following transformation of the generalized complex structure 
\beq 
{\cal J}_b = \left ( \begin{array}{ll}
                    1 & 0\\
                    b & 1
\end{array} \right) {\cal J} \left ( \begin{array}{ll}
                    \,\,\,\,\,1 & 0\\
                   -b & 1
\end{array} \right) 
\eeq{transofrmajdl}
 and thus gives rise to a new extended supersymmetry generator. 
 Therefore ${\cal J}_b$ is again the generalized complex
 structure. This is a physical explanation of the behavior of generalized complex 
 structure under $b$-transform.

Using $\delta_i (\epsilon) \bullet =\{ {\mathbf Q}_i(\epsilon), \bullet\}$ 
  we can write down the explicit form for the second
 supersymmetry transformations as follows
\beq
\delta_2(\epsilon) \Phi^\mu = \epsilon D\Phi^\nu J^\mu_{\,\,\nu} - \epsilon S_\nu P^{\mu\nu}
\eeq{trsnakdf;}
\beq
\delta_2(\epsilon) S_\mu = \epsilon D(S_\nu J^\nu_{\,\,\mu}) 
- \frac{1}{2} \epsilon S_\nu S_\rho P^{\nu\rho}_{\,\,\,\,,\mu}
 + \epsilon D(D\Phi^\nu L_{\mu\nu}) + \epsilon S_\nu D\Phi^\rho J^\nu_{\,\,\rho,\mu} -
 \frac{1}{2} \epsilon D\Phi^\nu D\Phi^\rho L_{\nu\rho,\mu}.
\eeq{tarsjS}
 Indeed it coincides with the supersymmetry transformation for the topological 
  model analyzed in \cite{Lindstrom:2004iw}\footnote{Namely, in \cite{Lindstrom:2004iw}  
  the transformations (4.2)-(4.3) subject to (4.5)-(4.6) coincide with (\ref{trsnakdf;})-(\ref{tarsjS})
  in the present paper, modulo obvious identifications.}.
   
Alternatively we can relate the generalized complex structure with an odd differential
  $\delta$ on $C^\infty(\Pi T^*{\cal L}M)$. Indeed the supersymmetry transformations (\ref{susycondhs})
 and (\ref{trsnakdf;})-(\ref{tarsjS}) can be thought of as odd transformations (by putting formally $\epsilon =1$)
 which squares to the translations, $\d$. Thus we can define the odd generator
\beq
{\mathbf q} = {\mathbf Q}_1(1) + i {\mathbf Q}_2(1) = - \int\limits_{S^{1,1}} d\sigma d\theta\,\,(S_\mu Q\Phi^\mu
 + i D\Phi^\rho S_\nu J^\nu_{\,\,\rho} + \frac{i}{2}D\Phi^\nu D\Phi^\rho L_{\nu\rho} + 
 \frac{i}{2} S_\nu S_\rho P^{\nu\rho}),
\eeq{nilpotenal}
 which gives rise to the following transformations
\beq
\delta \Phi^\mu = Q\Phi^\mu + iD\Phi^\nu J^\mu_{\,\,\nu} - i S_\nu P^{\mu\nu} ,
\eeq{nilp11}
\beq
\delta S_\mu = QS_\mu + i D(S_\nu J^\nu_{\,\,\mu}) 
- \frac{i}{2}  S_\nu S_\rho P^{\nu\rho}_{\,\,\,\,,\mu}
 + i D(D\Phi^\nu L_{\mu\nu}) + i S_\nu D\Phi^\rho J^\nu_{\,\,\rho,\mu} -
 \frac{i}{2}  D\Phi^\nu D\Phi^\rho L_{\nu\rho,\mu}. 
\eeq{nilp22}
 Thus $\delta^2 =0$ if and only if ${\cal J}$ defined in (\ref{definstardo}) is a generalized
 complex structure. In doing the calculations one should remember that now 
 $\delta$ is odd operation and whenever it passes through an odd object (e.g., $D$, $Q$ and
 $S$) there is extra minus. The existence of odd nilpotent operation (\ref{nilp11})-(\ref{nilp22}) is 
 related to the topological twist of $N=2$ algebra (for the related discussion see 
 \cite{Kapustin:2003sg} and \cite{Kapustin:2004gv}).
  The odd generator (\ref{nilpotenal}) is
 reminiscent of the solution of master equations proposed in \cite{Zucchini:2004ta}
  (see also \cite{Zucchini:2005rh}). However there are
  principal  differences related to the setup and to the definitions of basic 
 operations (e.g., $D$).

It is straightforward to generalize all results to the case when $H\neq 0$.
 In all formulas we can generate $H$ by non-canonical transformations
 \beq
  \Phi^\mu\,\,\rightarrow\,\,\Phi^\mu,\,\,\,\,\,\,\,\,\,\,\,\,\,\,\,\,\,\,\,
 S_\mu\,\,\rightarrow\,\,S_\mu + B_{\mu\nu} D\Phi^\nu
 \eeq{noncamsyah}
   with $H_{\mu\nu\rho} = B_{\mu\nu,\rho} + B_{\nu\rho,\mu} + B_{\rho\mu,\nu}$
 with $B_{\mu\nu,\rho} \equiv \d_\rho B_{\mu\nu}$. The transformation (\ref{noncamsyah}) is just a technical trick
 and all final formulas contain only $H$.
Thus the generator of 
  manifest supersymmetry is
\beq
 {\mathbf Q}_1(\epsilon) = - \int\limits_{S^{1,1}} d\sigma d\theta\,\, \epsilon (S_\mu +
    B_{\mu\nu} D\Phi^\nu)Q \Phi^\mu = \int\limits_{S^1} d\sigma \epsilon ( p_\mu \lambda^\mu -
     \rho_\mu \d X^\mu - \frac{1}{3} H_{\mu\nu\rho} \lambda^\mu\lambda^\nu \lambda^\rho)
\eeq{Htwistsusy1}
 and the generator of translations is
\beq
 {\mathbf P}(a) = \int\limits_{S^{1,1}} d\sigma d\theta\,\, a (S_\mu \d \Phi^\mu
 - \frac{1}{6} H_{\mu\nu\rho} D\Phi^\mu D\Phi^\nu  D\Phi^\rho).
\eeq{Htwisttran}
 Assuming the full symplectic structure (\ref{sympsuper}), 
  ${\mathbf Q}_1(\epsilon)$ and ${\mathbf P}(a)$  obey the same algebra (\ref{susyalg1}) 
  as before. The ansatz for the generator ${\mathbf Q}_2(\epsilon)$ of the second supersymmetry
    is the same as before (\ref{definchat}) and the algebra (\ref{susy2regwj})
      should be imposed.  In its turn the algebra implies the following conditions
\beq 
{\cal J}^2=-1_{2d},\,\,\,\,\,\,\,\,\,\,\,
 \Pi_{\mp} [\Pi_{\pm}(X+\eta), \Pi_{\pm}(Y+\eta)]_H=0,
 \eeq{twistgendkal}
 where $[\,\,.\,\,]$ is twisted Courant bracket. Therefore now ${\cal J}$ is a twisted generalized 
  complex structure.  

 There is a possibility to modify the supersymmetry algebra slightly \cite{Hull:1997kk}.
  Namely for the  "pseudo-supersymmetry" algebra 
     the last condition in (\ref{susy2regwj}) is replaced  by the following one
\beq
 \{ {\mathbf Q}_2(\epsilon), {\mathbf Q}_2(\tilde{\epsilon})\} = 
 - {\mathbf P}(2\epsilon\tilde{\epsilon}).
\eeq{pesudosusy}
 Geometrically it implies that 
\beq 
{\cal J}^2= 1_{2d},\,\,\,\,\,\,\,\,\,\,\,
 \Pi_{\mp} [\Pi_{\pm}(X+\eta), \Pi_{\pm}(Y+\eta)]_H=0,
 \eeq{prodsteh}
 where $\Pi_\pm=\frac{1}{2}(1_{2d} \pm  {\cal J})$. Using the fact ${\cal J}$ respects
  the natural pairing we conclude that  $\Pi_\pm$ project to maximally isotropic 
   subbundles which are involutive with respect to (twisted) Courant bracket.
 Therefore we get two complementary (twisted) Dirac structures $L_+$ and  $L_{-}$
 such that  $T\oplus T^* = L_+ \oplus L_-$. For any $M$ there is always a trivial 
 ``pseudo-supersymmetry''
\beq
\delta_2(\epsilon) \Phi^\mu = \epsilon D\Phi^\mu,\,\,\,\,\,\,\,\,\,\,\,\,\,\,\,\,\,\,\,
\delta_2(\epsilon) S_\mu = - \epsilon DS_\mu,
\eeq{pesudosusy1ex}
 which corresponds to the choice ${\cal J}= 1_{2d}$. Another interesting example of 
 ``pseudo-supersymmetry'' is given by the following choice
\beq
 {\cal J}= \left (\begin{array}{ll}
                 1_d & P \\
                 0 & -1_d \end{array} \right ),
\eeq{poissonaexap}
 where $P$ is a Poisson structure on $M$. In this case the transformations are 
\beq
\delta_2(\epsilon) \Phi^\mu = \epsilon D\Phi^\mu - \epsilon S_\nu P^{\mu\nu},\,\,\,\,\,\,\,\,\,\,
\delta_2(\epsilon) S_\mu =  - \epsilon DS_\mu 
- \frac{1}{2} \epsilon S_\nu S_\rho P^{\nu\rho}_{\,\,\,\,,\mu}.
\eeq{pesudosusy2ex}
 In analogy with the discussion of the standard $N=2$ supersymmetry we can consider the topological
  twist of ``pseudo-supersymmetry''. Namely we can introduce an odd nilpotent operation 
  on $\Pi T^*{\cal L}M$
 as follows $\delta = \delta_1(1) + \delta_2(1)$. Thus for the example (\ref{poissonaexap}) the 
 corresponding nilpotent operation $\delta$ is reminiscent of the BV-transformations of 
 the Poisson sigma model \cite{Cattaneo:1999fm} (however we are working in Hamiltonian setup). 

The above discussion about the odd transformations can be summarized as follows
\begin{proposition}
  The super-Poisson algebra $C^\infty(\Pi T^* {\cal L}M)$ admits an odd differrential 
  $\delta$ if either $M$ is generalized complex manifold or $M$ is 
   Dirac manifold with two complementary Dirac structures.
\end{proposition}

\section{D-branes}
\label{s:branes}

 We can generalize the previous discussion to the world-sheets with the boundary. 
 For the world-sheet $\Sigma = P^1 \times {\mathbb R}$ with $P^1$ being the interval $[0,1]$
 the phase space  can be identified with the cotangent bundle $T^*PM$ of the path space
 $PM = \{ X: [0,1] \rightarrow M,\,X(0) \in D_0,\,X(1) \in D_1\} $ where $D_0$ and 
 $D_1$ are submanifolds of $M$. To write down the symplectic structure on $T^*PM$
 we have to require that $D_0$ and $D_1$ are generalized submanifolds of $M$, 
 i.e. there exists $B^i \in \Omega^2(D^i)$ such that $dB^i = H|_{D_i}$. Hence the symplectic 
 structure is given by
$$ \omega = \int\limits^{1}_{0} d\sigma\,\,(\delta X^\mu \wedge \delta p_\mu + \frac{1}{2} H_{\mu\nu\rho} \d X^\mu 
 \delta X^\nu \wedge \delta X^\rho ) +$$
\beq
+ \frac{1}{2} B^0_{\mu\nu}(X(0)) \delta X^\mu(0) \wedge \delta X^\nu(0) -
 \frac{1}{2} B^1_{\mu\nu}(X(1)) \delta X^\mu(1) \wedge \delta X^\nu(1),
\eeq{defsympboun}
 where the boundary contributions are needed in order $\omega$ to be closed.
 
  Next we proceed with the supersymmetrization of $T^*PM$. In analogy with the previous 
   discussion we introduce 
   the "superinterval" $P^{1,1}$  with coordinates $(\sigma, \theta)$ and the superinterval space
    ${\cal P}M = \{ \Phi: P^{1,1} \rightarrow M\}$.
   The phase space is given by the cotangent bundle $\Pi T^* {\cal P}M$ of ${\cal P}M$,
    with reversed parity on the fibers. 
    As before we introduce two superfields $\Phi$ and $S$, see (\ref{superfiel}). Let us start 
     to discuss the situation when $H=0$. The canonical symplectic structure on $\Pi T^*{\cal P}M$  
 is given by 
 \beq
  \omega = \int\limits_{P^{1,1}} d\sigma d\theta\,\,\delta \Phi^\mu \wedge \delta S_\mu .
 \eeq{canonsympbo}
 However this symplectic form is not supersymmetric unless the extra data is specified. 
  In presence of boundaries the superfield expressions are not automatically supersymmetric
   due to possible boundary terms. Namely the transformation of symplectic form (\ref{canonsympbo})
   under manifest supersymmetry (\ref{susycondhs}) gives rise to a boundary term
   \beq
    \delta_1 (\epsilon) \omega = \epsilon (\delta X^\mu (1)\wedge \delta \rho_\mu(1)
     - \delta X^\mu(0) \wedge \delta \rho_\mu(0)).  
   \eeq{boundtermsym} 
   Moreover the supersymmetry algebra has boundary term
   \beq
     \{ {\mathbf Q}_1(\epsilon) , {\mathbf Q}_1(\tilde{\epsilon})\} =
      P(2\epsilon \tilde{\epsilon}) + 2(\rho_\mu(1) \lambda^\mu(1) -\rho_\mu(0)\lambda^\mu(0))
   \eeq{susymanbound}
   and ${\mathbf Q}_1(\epsilon)$ is not translation-invariant
   \beq
    \{ {\mathbf P}(a), {\mathbf Q}_1(\epsilon) \} = a\epsilon (p_\mu(1) \lambda^\mu(1)
    - p_\mu(0) \lambda^\mu(0)  + \rho_\mu(0) \d X^\mu(0) - \rho_\mu(1)\d X^\mu(1)) .
   \eeq{extrabdpq}
 These boundary terms spoil wanted properties. The problem can be cured by 
  imposing the appropriate boundary conditions on the fields such that the boundary 
   terms vanish. In fact the required boundary conditions have a simple geometrical form
   \beq
    \Lambda(1) =  \left ( \begin{array}{l}
                    D\Phi(1)\\
                      S(1)
\end{array} \right) \in X^*(\Pi(TD_1 \oplus N^*D_1)),\,\,\,\,\,\,\,\,\,\,\,\,\,\,\,
  \Lambda(0) =  \left ( \begin{array}{l}
                    D\Phi(0)\\
                      S(0)
\end{array} \right) \in X^*(\Pi(TD_0 \oplus N^*D_0))
   \eeq{boundarycondsp}
 where $TD_i$ is tangent and $N^* D_i$ is conormal bundles of $D_i$ correspondingly.
 The conditions (\ref{boundarycondsp}) are understood as conditions on each component of superfields. 

 Next we consider the case when $H\neq 0$.  We should apply the same logic as before.
  Namely we have to impose such boundary conditions on the fields that there are 
   no unwanted boundary terms which spoil supersymmetry. 
  Consequently we arrive to the following symplectic form for $\Pi T^*{\cal P}M$
$$ \omega = \int\limits_{P^{1,1}} d\sigma d\theta\,\,(\delta \Phi^\mu \wedge \delta S_\mu -
 \frac{1}{2} H_{\mu\nu\rho} D\Phi^\mu \delta \Phi^\nu \wedge \delta \Phi^\rho ) +$$
\beq
+ \frac{1}{2} B^0_{\mu\nu}(X(0)) \delta X^\mu(0) \wedge \delta X^\nu(0) -
 \frac{1}{2} B^1_{\mu\nu}(X(1)) \delta X^\mu(1) \wedge \delta X^\nu(1),
\eeq{defsymboususy}
 with the fields satisfying the following boundary conditions
\beq
 \Lambda(1) \in X^*(\Pi \tau_{D_1}^{B^1}),\,\,\,\,\,\,\,\,\,\,\,\,\,\,\,\,\,\,\,\,\,
  \Lambda(0) \in X^*(\Pi \tau_{D_0}^{B^0}),
\eeq{defboundgenal}
 where $\tau_{D_i}^{B^i}$ is a generalized tangent bundle (\ref{deftbundlsub}) 
 of generalized submanifold $(D_i, B^i)$.  With these boundary conditions 
  the symplectic form (\ref{defsymboususy}) is supersymmetric and there are no boundary terms
   in the supersymmetry algebra.
  Actually the boundary conditions
  (\ref{defboundgenal}) can be thought of as $B$-transform of the conditions
   (\ref{boundarycondsp}). The spaces $\tau_{D_i}^{B^i}$ are maximally  isotropic with 
    respect to the natural pairing $\langle \,\,,\,\,\rangle$ on $T\oplus T^*$, i.e.
  \beq
   \langle \Lambda(0), \Lambda(0) \rangle =0,\,\,\,\,\,\,\,\,\,\,\,\,\,\,\,\,\,
   \langle \Lambda(1), \Lambda(1) \rangle =0.
  \eeq{isotrocon}
Finally we have constructed the the supersymmetric version of $\Pi T^*{\cal P}M$
 where the boundary conditions (\ref{defboundgenal}) play the crucial role. 

Now we turn to the discussion of extended supersymmetry. As in previous section 
 we should write down the generator for second supersymmetry (\ref{definchat})
 and check the algebra (\ref{susy2regwj}). 
  The only difference with the discussion from section \ref{s:susy} is that we should keep 
  track of the boundary terms. For example, we can check if ${\mathbf Q}_2(\epsilon)$
   is translation-invariant, i.e. 
\beq
 \{ {\mathbf P}(a) , {\mathbf Q}_2(\epsilon) \} = \frac{1}{2} a \int d\theta\,\, \epsilon \langle
 \Lambda, {\cal J} \Lambda \rangle (0) - \frac{1}{2} a \int d\theta\,\, \epsilon \langle
 \Lambda, {\cal J} \Lambda \rangle (1).
\eeq{commuta}
 Thus translation-invariance is spoiled by the boundary terms. We can restore it by 
  imposing the additional property
  \beq
   \langle
 \Lambda, {\cal J} \Lambda \rangle (0)=0,\,\,\,\,\,\,\,\,\,\,\,\,\,\,\,\,\,\,\,
 \langle
 \Lambda, {\cal J} \Lambda \rangle (1)=0 .
  \eeq{aditspmjdkl}
 Indeed this property (together with (\ref{defboundgenal})) is sufficient to cancel
   all other unwanted boundary terms, e.g. 
  in (\ref{susy2regwj}) or in $\delta_2(\epsilon)\omega$. Thus the property (\ref{aditspmjdkl})
  together with (\ref{defboundgenal}) implies that the subbundles $\tau_{D_i}^{B^i}$ are stable under
  ${\cal J}$, i.e. $(D_i, B^i)$ are the generalized complex manifolds introduced 
   in \cite{Gualtieri}.
 
We summarize the above discussion in proposition
\begin{proposition} 
The cotangent bundle $\Pi T^*{\cal P}M$ admits
 $N=2$ supersymmetry if and only if $M$ is (twisted) generalized complex manifold 
  and $(D_i, B_i)$ are generalized complex submanifolds.  
 \end{proposition}
 This result agrees with the previous considerations in \cite{Lindstrom:2002jb} and
  \cite{Zabzine:2004dp}. However 
 now it is applicable to a wide class of sigma models. 

If instead we consider the 
 ``pseudo-supersymmetry'' algebra (\ref{pesudosusy}) then appropriate boundary conditions would 
 require that $\tau_{D_i}^{B^i}$ is invariant under ${\cal J}$ which defines two transversal 
 Dirac structures. This is a real analog of the generalized complex submanifold. 
 In the example (\ref{poissonaexap}), $\tau_{D}^{0}$ is stable under ${\cal J}$ if 
  a submanifold $D$ is coisotropic with respect to $P$, see \cite{Cattaneo:2003dp}.

\section{Concluding remarks}
\label{s:end}

In this short note we clarified and extended results from \cite{Lindstrom:2004iw}. 
 The first order actions discussed in \cite{Lindstrom:2004eh} and 
  \cite{Lindstrom:2004iw} can be thought of as phase space actions and therefore 
 the Hamiltonian formalism should naturally arise in the problem.
  Indeed the Hamiltonian formalism offers a deep insight on the relation 
 between the world-sheet and the geometry of $T\oplus T^*$. 
 The main result of the  paper is that
 the  phase space realization of extended supersymmetry 
 is related to generalized complex structure. This result is model independent 
 and it is applicable to the wide range of sigma models, e.g. the standard sigma model, 
 the Poisson sigma model, the twisted Poisson sigma model etc. 

The next step would be to specify the Hamiltonian ${\cal H}$ (i.e., choose
 the concrete model) and check that ${\mathbf Q}_2$ is in fact the symmetry of 
 the Hamiltonian. At this stage the compatibility between the geometrical data 
 used in ${\cal H}$ (e.g., a metric $g$, a Poisson structure $\pi$ etc.)
 and ${\cal J}$ will arise. We hope to come back to this elsewhere. 

\bigskip

\bigskip

{\bf Acknowledgements}:
 I am grateful to Anton Alekseev, Alberto Cattaneo, Giovanni Felder, 
 Matthias Gaberdiel and Ulf Lindstr\"om  for discussions on this
  and related subjects. I thank Ulf Lindstr\"om and Pierre Vanhove 
 for reading and commenting on the manuscript. 
The research is  supported by EU-grant MEIF-CT-2004-500267.

\appendix
\Section{Appendix: superspace conventions}

We use the superspace conventions. The odd coordinate is labeled by $\theta$
 and the covariant derivative $D$ and supersymmetry generator $Q$ are defined as follows
\beq
 D = \d_\theta - \theta \d_\sigma,\,\,\,\,\,\,\,\,\,\,\,\,\,\,
 Q=  \d_\theta + \theta \d_\sigma
\eeq{relaon1d}
 such that 
\beq
 D^2 = - \d_\sigma,\,\,\,\,\,\,\,\,\,\,\,
 Q^2 = \d_\sigma,\,\,\,\,\,\,\,\,\,\,\,
 QD +DQ =0 .
\eeq{properDQ}
 In terms of covariant derivatives, a supersymmetry transformation\footnote{We give the expressions for the case 
 $H=0$. Analogously using the generator (\ref{Htwistsusy1}) one can write down the expressions for the case $H \neq 0$.} 
 of a superfields is then given by
\beq
 \delta_1(\epsilon) \Phi^\mu \equiv  \epsilon Q\Phi^\mu,\,\,\,\,\,\,\,\,\,\,\,\,\,\,\,\,\,
 \delta_1(\epsilon) S_\mu \equiv  \epsilon Q S_\mu .
\eeq{susycondhs}
The components of superfields can be found via projection as follows,
\beq
 \Phi|\equiv X^\mu,\,\,\,\,\,\,\,\,\,\,\,
 D\Phi| \equiv \lambda^\mu,\,\,\,\,\,\,\,\,\,\,\,
 S_\mu|\equiv \rho_\mu,\,\,\,\,\,\,\,\,\,\,\,
 DS_\mu| \equiv p_\mu,
\eeq{projsupers}
 where a vertical bar denotes ``the $\theta=0$ part of''. Thus, in components, 
 the N=1 supersymmetry transformations are given by 
\beq
 \delta_1(\epsilon) X^\mu = \epsilon \lambda^\mu,\,\,\,\,\,\,\,\,
 \delta_1(\epsilon) \lambda^\mu = - \epsilon \d X^\mu, \,\,\,\,\,\,\,\,
 \delta_1(\epsilon) \rho_\mu =\epsilon p_\mu,\,\,\,\,\,\,\,\,
 \delta_1(\epsilon) p_\mu = - \epsilon \d \rho_\mu.
\eeq{manisusycomp}
The N=1 spinorial measure in terms of covariant 
 derivatives 
\beq
 \int d\theta\,\, {\cal L} = D {\cal L}|
\eeq{measurein}

\Section{Appendix: basics on $T \oplus T^*$}

Consider the vector bundle $T \oplus T^*$ which is the sum of the tangent and cotangent 
 bundles of an $d$-dimensional manifold $M$. $T\oplus T^*$ has a natural pairing
\beq
  \langle X + \xi, Y + \eta \rangle \equiv (i_Y \xi + i_X \eta) \equiv
 \left (\begin{array}{l}
 X\\
 \xi \end{array} \right )^t {\cal I}
\left (\begin{array}{l}
 Y\\
 \eta \end{array}\right ) .
\eeq{definnaturka}
 The smooth sections of $T\oplus T^*$ have a natural bracket operation called
  the Courant bracket and defined as follows
\beq
 [X+ \xi, Y + \eta]_c = [X, Y] + {\cal L}_X \eta - {\cal L}_Y \xi  -\frac{1}{2}d (i_X\eta - i_Y \xi)
\eeq{defCourant}
 where $[\,\,,\,\,]$ is a Lie bracket on $T$. Given a closed three form $H$ we can define
  a twisted Courant bracket
\beq
 [ X+ \xi, Y + \eta]_H = [X+\xi, Y+\eta ]_c + i_X i_Y H .
\eeq{deftwCour}
 The orthogonal automorphism (i.e., such which preserves $\langle\,\,,\,\,\rangle$)
   $F:T\oplus T^* \rightarrow \rightarrow T\oplus T^*$ of (twisted)
 Courant bracket 
\beq
 F([X+\xi, Y + \eta ]_H) = [F(X+\xi), F(Y+\eta)]_H
\eeq{automspaorp}
 is semidirect product of $Diff(M)$ and $\Omega^2_{closed}(M)$,
 where the action of the closed two form is given as follows
\beq
 e^b(X+\xi) \equiv X + \xi + i_X b
\eeq{actionwtalk}
 for $b\in \Omega^2_{closed}(M)$. The transformation (\ref{actionwtalk}) is called $b$-transform.
 The maximally isotropic subbundle $L$ of $T\oplus T^*$, which is involutive with respect 
 to (twisted) Courant bracket is called (twisted) Dirac structure. We can consider two complementary 
 (twisted) Dirac structures $L_+$ and $L_-$ such that $T\oplus T^* = L_+ \oplus L_-$. Alternatively 
 we can define $L_\pm$ by proving a map ${\cal J}:T\oplus T^* \rightarrow T\oplus T^*$ with 
 the following properties
\beq
 {\cal J}^t {\cal I} = - {\cal I} {\cal J},\,\,\,\,\,\,\,\,
 {\cal J}^2 = 1_{2d},\,\,\,\,\,\,\,\,
 \Pi_{\mp} [\Pi_{\pm} (X+\xi), \Pi_{\pm}(Y+\eta)]_H =0
\eeq{transverds}
 where $\Pi_\pm=\frac{1}{2} (1_{2d} \pm {\cal J})$ are projectors on $L_\pm$.

 The (twisted) generalized complex structure is the complex version of two complementary (twisted) 
 Dirac subbundles
 such that  $(T\oplus T^*)\oplus {\mathbb C} = L \oplus \bar{L}$. We can define the generalized 
 complex structure as a map ${\cal J}: (T\oplus T^*)\otimes {\mathbb C} 
  \rightarrow (T\oplus T^*)\otimes {\mathbb C}$ with 
 the following properties
\beq
 {\cal J}^t {\cal I} = - {\cal I} {\cal J},\,\,\,\,\,\,\,\,
 {\cal J}^2 = - 1_{2d},\,\,\,\,\,\,\,\,
 \Pi_{\mp} [\Pi_{\pm} (X+\xi), \Pi_{\pm}(Y+\eta)]_H =0,
\eeq{transverds}
 where $\Pi_\pm=\frac{1}{2} (1_{2d} \pm i{\cal J})$ are the projectors on $L$ and $\bar{L}$ correspondingly. 

 The generalized submanifold is a sumanifold $D$ with a two form $B \in \Omega^2(M)$ such 
 that $dB = H|_D$. For generalized submanifold $(D,B)$ we can define the generalized
 tangent bundle 
\beq
\tau_D^B = \{ X +\xi \in T D\oplus T^*M|_D,\,\,\, \xi|_D = i_X B\}.
\eeq{deftbundlsub}
 The submanifold $(D,B)$ is (twisted) generalized complex submanifold if $\tau_D^B$ is stable under
 the action of map ${\cal J}$ defined in (\ref{transverds}).

For further details the reader may consult the Gualtieri's thesis \cite{Gualtieri}.

\end{document}